# Ultrafast ring opening in CHD investigated by simplex-based spectral unmixing

*J. L. White[1,2], J. Kim[1,3], V. S. Petrović[1,3], P. H. Bucksbaum[1,2,3]*
[1]*PULSE Institute,* [2]*Department of Applied Physics, and* [3]*Department of Physics, Stanford University, Stanford, CA 94305*

We use spectral unmixing to determine the number of transient photoproducts and to track their evolution following the photo-excitation of 1,3-cyclohexadiene (CHD) to form 1,3,5-hexatriene (HT) in the gas phase. The ring opening is initiated with a 266 nm ultraviolet laser pulse and probed via fragmentation with a delayed intense infrared 800 nm laser pulse. The ion time-of-flight (TOF) spectra are analyzed with a simplex-based spectral unmixing technique. We find that at least three independent spectra are needed to model the transient TOF spectra. Guided by mathematical and physical constraints, we decompose the transient TOF spectra into three spectra associated with the presence of CHD, $CHD^+$, and HT, and show how these three products appear at different times during the ring opening.

## I. Introduction

The ring opening of 1,3-cyclohexadiene (CHD) to form hexatriene (HT) belongs to a class of pericyclic reactions [1] found in many biochemical pathways, the most important of which is the biosynthesis of vitamin $D_3$ from its provitamin dehydrocholesterol. For this reason, CHD has been a model for studies of non-radiative photoisomerization.

The current understanding of CHD-HT isomerization comes from time-resolved multi-photon fragmentation experiments with mass-resolved detection [2-9], REMPI spectroscopy [10], resonance Raman excitation spectroscopy [11], electron diffraction [12], electron energy loss measurements [13], and liquid phase transient absorption measurements [14-16]. Experiments that control the CHD:HT branching ratio have been performed in both gas phase [2] and liquid phase [17, 18]. The isomerization has also been extensively studied computationally [19-22]. It is now generally accepted that the photoinitiated ring opening is governed by non-adiabatic interactions; at least two conical intersections are found to play a role in the photoinitiated isomerization (Fig. 1). When an ultraviolet pulse launches a wavepacket from the $S_0$ ground state of CHD onto the first excited $S_1$ state, most of the wavepacket passes through an $S_1/S_2$ conical intersection onto the lower branch after approximately 55 fs [9]. The wavepacket then accelerates towards another conical intersection as shown between the S$_1$ and S$_0$ surfaces where it bifurcates and continues simultaneously towards both the HT and CHD ground states. The isomerization to the cZc conformation of HT is complete in approximately 200 fs [9]. Theory suggests that the CHD:HT branching ratio is approximately 50:50 for isolated molecules [19]; a CHD:HT ratio of 60:40 was reported for experiments in solution phase [23].

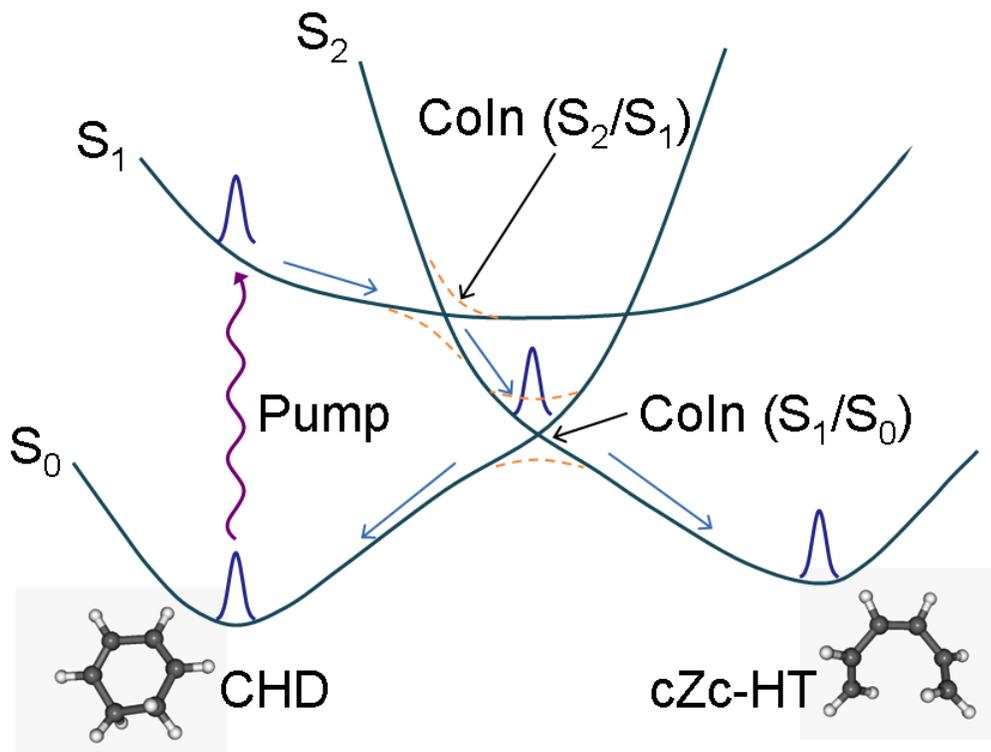

Figure 1: Cartoon of CHD and HT potential energy surfaces along the reaction coordinate.

Differences in the laser-induced photo-fragmentation patterns of CHD and HT have been used to track the time evolution of the conversion to HT following photo-excitation of CHD. Short intense 800 nm laser pulses have been shown to lead to easily distinguishable fragment patterns for these two isomers [7]. The difference in the fragmentation patterns has been attributed to differences in photoabsorption in the cations [7, 24, 25]. Kotur et al. used this observation to guide a control experiment in which they were able to increase the HT yield by shaping the UV launch pulse [2]. Fuss et al. reported a time-resolved laser fragmentation experiment in which they fit the ion signals to exponentials in order to relate the temporal evolution of certain fragments to population passing through specific locations along the isomerization pathway [7]. Fuss et al. note that each time constant appears in several mass peaks, but ultimately conclude that monitoring masses 79 and 80 (corresponding to $C_6H_7$ and $C_6H_8$) is adequate in order to follow the isomerization. Kosma et al. monitor masses 79 and 80 when performing a similar experiment with 13 fs resolution [9].

In the work of Fuss et al., Kotur et al., and Kosma et al., typically a pair of time-of-flight mass peaks is used to follow the isomerization between the two molecules. Decrease in the parent peak ion count (i.e., mass 80) is associated with the disappearance of the CHD isomer, while the increase of the H ion count is associated with the formation of HT, as is the increase in the count of ion fragments with two C atoms. In all these examples, the choice of fragments is based on empirical evidence. When only two fragments are used to follow the isomerization, one can discriminate between at most two species (unless other parameters are also observed, such as time evolution). Here we will use the term *species* to refer not only to chemical species, but to any molecular or ionic entity resulting in a distinctive fragmentation pattern. If more species are present (perhaps in the form of transient spectral features or unanticipated reaction products) a two-mass peak analysis must underfit the data. More information contained in other peaks in the mass spectra could be used.

In this paper we demonstrate an analysis based on geometric spectral unmixing to determine the number of fragmentation patterns and estimate their TOF spectra without prior knowledge of the number of species present [26]. The method benefits from the richness and redundancy of information contained in the *complete* time-of-flight fragmentation mass spectrum and not only a subset of peaks. This procedure is especially useful in cases where one (or more) of the species involved is transient or otherwise not available for direct study. We apply this spectral unmixing technique to a time-resolved study of CHD-HT isomerization and find that the fragmentation patterns of *three species* compose the TOF spectra collected during the UV-initiated isomerization of CHD. We assign the species to CHD, cZc-HT, and $CHD^+$ and estimate their laser fragmentation time-of-flight mass spectra. We then estimate their proportions in the fragmentation mass spectrum as a function of probe delay and discuss the time dependence of the underlying processes.

## II. Experiment

Experiments were performed in an effusive beam of 1,3-cyclohexadiene (Aldrich, 97%, no further purification), skimmed and expanded at room temperature (its vapor pressure is approximately 10 mbar [27]). A commercial Ti:sapphire laser produced 2.5 mJ pulses with 80 fs FWHM duration centered at 800 nm wavelength at 1 kHz repetition rate. A 100 µJ portion of this beam (referred to below as the IR beam) was delayed by -400 fs to +600 fs (20 fs steps) and used as the fragmentation pulse. Another portion of the Ti:sapphire beam (600 µJ) was used to create the UV excitation pulse by Type I frequency doubling in BBO (400 µm) and subsequent Type I frequency summing of 400 nm and 800 nm pulses in BBO (300 µm). The resulting UV pulse had <30 µJ pulse energy and ~120 fs duration FWHM centered at 266 nm. The UV pulse energy was set between the multiphoton ionization and fragmentation thresholds for CHD.

The IR pulse energy was chosen such that when CHD was fragmented, many different mass fragments were present but not every CHD molecule was being ionized or fragmented. Having many different fragments is advantageous when unmixing spectra. Figure 2 shows how the TOF spectrum of CHD varies as a function of the IR pulse energy (0~120 µJ) in an IR-only experiment. For pump-probe experiments, the IR pulse energy was ~100 µJ.

In pump-probe experiments, the two laser beams propagated collinearly through the chamber and intersected the gas jet at 90°. Overlap of the foci in the direction of propagation was achieved by detuning a telescope in the IR beam path. At the focus, the UV beam was larger than the IR beam. In the interaction region, the UV and IR beam intensities can be estimated by the ionization threshold as well as the geometry of the beams and other similar factors. We estimate the UV beam intensity was ~$10^{11}$ W/cm$^2$ and the IR beam intensity ~$10^{13}$ W/cm$^2$.

Ions were extracted with a 750 V/cm external field (always present) in the direction perpendicular to both the laser beams and molecular beam and collected at a multichannel plate detector after propagating 11 cm. A Poisson statistical analysis of the signal leads to an estimate that several hundred molecules were dissociated or ionized on each laser pulse, the majority of them in the parent ion group. The strong IR field and the large number of molecules fragmented on each shot mean that each fragmentation spectrum is the result of summing over all possible dissociation/ionization channels. The UV pump pulse alone was responsible for approximately half of the parent ions.

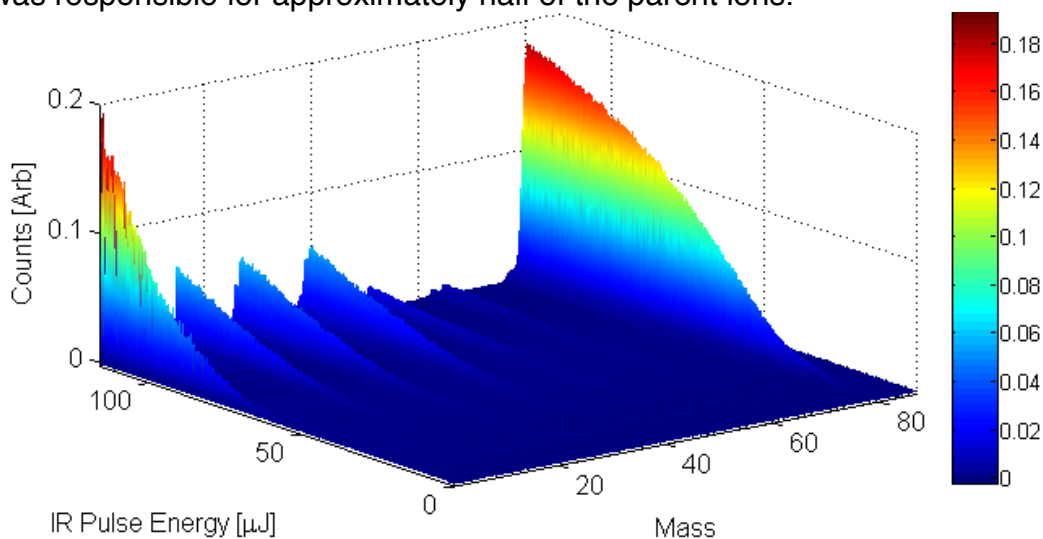

Figure 2: CHD TOF mass spectrum as a function of IR pulse energy in an IR-only experiment. For pump-probe experiments, the IR pulse energy was chosen such that multiple daughter ions were present.

### III. Unmixing spectra

The TOF spectral data in our experiment are the result of various processes, such as: (1) UV-only production of $CHD^+$; (2) fragmentation of CHD, $CHD^+$ or HT by the IR probe laser; or (3) fragmentation of transient intermediates. Our analysis addresses several questions: (1) How many separate physical species or processes contribute to the ion spectra? (2) Does each have a characteristic TOF spectral signature? (3) What is the underlying physics?

Experimental spectral data are often mixtures of a small set of pure spectra from several distinct species, or *endmembers*. Many statistical analysis techniques have been developed to *unmix* a data set containing a mixture of distinct species to yield pure spectra, called *endmember spectra*, under a variety of assumptions, including cases where the number and identities of the endmember spectra are initially unknown. A review of spectral unmixing from a remote sensing perspective is given in reference [26] and from a chemometrics perspective in reference [28] (where it is referred to as multivariate curve resolution). Here we follow an intuitive geometric unmixing method known in both fields [29, 30]. We determine the number of endmembers needed to model the experimental TOF spectra, estimate the endmember spectra (using both physical and geometric knowledge), and finally unmix the experimental spectra to determine their composition as a function of pump-probe delay.

The key intuition underlying this approach is that the experimental spectra will be contained in a simplex whose vertices are the pure spectra. An (n-1)-simplex is an (n-1)-dimensional analog of a triangle or tetrahedron and has n vertices. It is the smallest convex set containing the vertices. A key property of simplices is that any point in the simplex can be written as a linear combination of the vertices in which the coefficients are nonnegative and sum to one.

An experimental spectrum with m spectral channels can be considered a vector in an m-dimensional vector space. We expect that the experimental spectra will be linear combinations of just a few (n << m) endmember spectra and noise. Using this linear mixture model, each experimental spectrum $S_{exp}$ can be written

$$\mathbf{S}_{exp} = \Sigma\, c_i\, \mathbf{E}_i + \varepsilon \qquad (1)$$

where the $E_i$ are the (possibly unknown) endmember spectra, $c_i$ is the fraction of the $i^{th}$ endmember spectrum found in $S_{exp}$, and $\varepsilon$ is noise. We require $c_i \geq 0$ on physical grounds. Since the solution to Equation (1) is not unique when the $E_i$ are unknown, one can further require that $\Sigma\, c_i = 1$ to facilitate the interpretation of the $c_i$ as proportions of the endmember spectra $E_i$. It is always possible to force the proportions $c_i$ to sum to one by choosing vertices $E_i$ that enclose the data. We then have

$$\mathbf{S}_{exp} = \sum c_i \mathbf{E}_i + \varepsilon, \quad c_i \geq 0, \quad \sum c_i = 1 \qquad (2)$$

Equation (2) is the convex geometry model (CGM), and it is equivalent to requiring that the data (neglecting noise) lie inside a simplex whose vertices are the endmember spectra $E_i$. Sometimes one of the vertices corresponds to a constant (possibly all zero) background; this background endmember is called the *shade endmember* (from its use to account for lighting differences in remote sensing applications) and its coefficient expresses the relative abundance of the constant background in a spectrum. In chemometrics it is common to divide each spectrum by its size (the sum of the counts in the spectrum, also called the 1-norm) in order to map spectra with the same shape to the same point; this scaling has the special property that it preserves the CGM and equation (2) still holds although the $c_i$ will be different [29]. We choose not to scale our spectra for a number of reasons. The most important reason is that differences in spectrum size (1-norm) contain physically relevant information. Spectrum size reflects the number of ions collected and so is a measure of the fragmentation cross section. Moreover, the presence of transient resonances would mean that the relative abundance of each species determined from scaled spectra would be wrong. Additionally, our collected spectra do not vary greatly in spectrum size, so little would be gained from scaling. The lack of variation is not surprising since, after averaging, the number of molecules in the interaction region and the laser pulse energies are both nearly constant. Finally, scaling the spectra is sensitive to the choice of constant background.

    We can estimate unknown endmember spectra by determining the vertices of a simplex enclosing the experimental spectra. The purer the spectra inside the simplex, the better the endmember estimation will be. Algorithms for automated endmember extraction exist [30-33], but we do not use them in the present work.

    To determine the number of endmembers, we first use a dimension reduction technique to determine the effective dimensionality of the data; here we use principal component analysis (PCA) but the partial least squares and minimum noise fraction [34] transforms are also widely used. PCA is a method for determining, within a high-dimensional space, the subspace that contains the data [35]. (More correctly, PCA determines an *affine* subspace, that is, a translation of a subspace.) This *p*-dimensional affine subspace contains the largest fraction of the variance of the data of any *p*-dimensional subspace, and projecting the original spectra into this affine subspace gives the best *p*-dimensional approximation to the original spectra in a least-squares sense. The affine subspace is found by diagonalizing the covariance matrix (of the spectra) and using the mean spectrum (the vector resulting from taking the arithmetic

mean of all the experimental spectra) plus the *p* eigenvectors with the largest eigenvalues (variances) to approximate the spectra. These *p* eigenvectors are called the *principal components* (PCs). The first PC is the direction of largest variance in the data, the second PC is the direction of second largest variance in the data (perpendicular to the first PC), and so on for subsequent PCs. The PCs form a basis for the affine space. The simplex containing the data will lie in this space.

If the PCs and the affine vector together span an n-dimensional subspace, then there are at most n endmembers (simplex vertices). In addition to helping to estimate the number of endmembers, dimension reduction also makes it easier to visualize the simplex (and how the spectra are mixed). Figure 3 shows the geometry of the affine subspace for our TOF spectra; in this case the data falls into a two-dimensional subspace (p=2) and the simplex can have at most 3 vertices (a triangle, n=3).

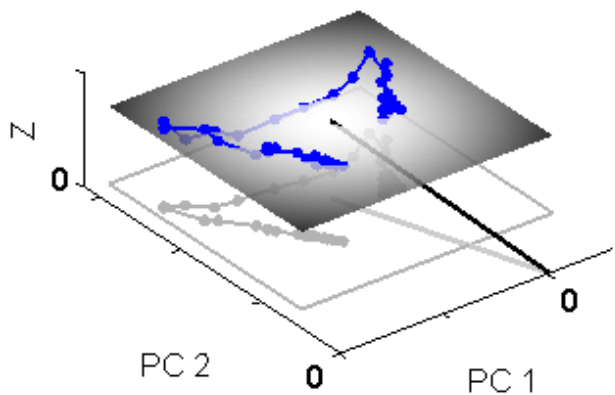

Figure 3: Geometry of the UV-800 pump-probe PC affine subspace containing the TOF spectra collected as CHD isomerizes. The spectra have been projected into the three dimensions determined by the first 2 PCs and the mean spectrum (black line). The Z direction is the component (normalized to have unit length) of the mean spectrum that is orthogonal to the first 2 PCs. A region in the PC space is indicated by the gray plane. Each experimental spectrum is shown as a blue dot, and blue lines connect adjacent delays (separated by 20 fs). Three spectra are required to model the data (for instance, the mean vector and two PCs, or three pure spectra). The projection onto the Z=0 plane is shown in light gray.

**IV. Results**

Figure 4 shows the TOF mass spectra as a function of pump-probe delay. Each mass spectrum is an average of 600 laser shots. The parent peak (mass 80) is always present but reaches a maximum at zero delay. The $H^+$ peak is almost absent at negative delays, and reaches its maximum around 180 fs delay, after which it decreases to a constant value.

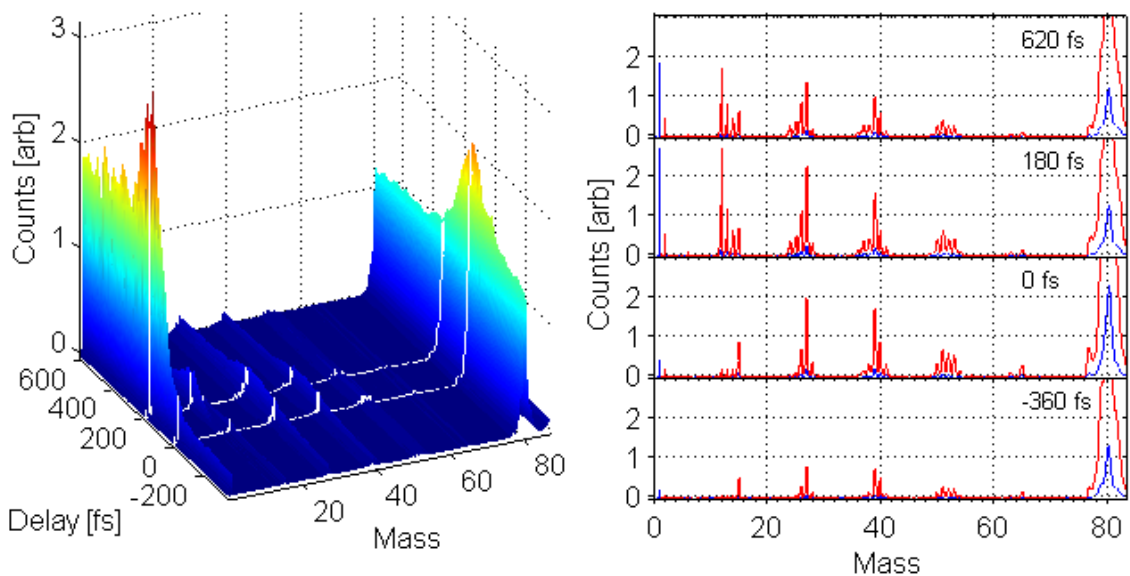

Figure 4: (left) TOF mass spectra as a function of pump-probe delay. Lines have been added to highlight the delays corresponding to the maximum yields of masses 1 and 80 (at 180 fs and 0 fs, respectively). (right) TOF mass spectra for several delays, including the earliest delay and those highlighted in white on the left-hand plot. The blue lines are the raw spectra and the red lines show the spectra magnified 8 times (the hydrogen peak is not shown magnified).

When PCA is applied to the collection of TOF mass spectra recorded at different delays in the pump-probe experiment, only two principal components account for more than 95% of the variance in the spectra (Fig. 5 inset). This implies that the collected TOF mass spectra very nearly lie in a two dimensional affine subspace of the detection space. This affine subspace does not contain the origin. Consequently three linearly independent vectors are needed to span that affine subspace (for example, the mean vector and the two most significant PCs, as shown in figure 3), suggesting that three endmember spectra are needed to explain the data.

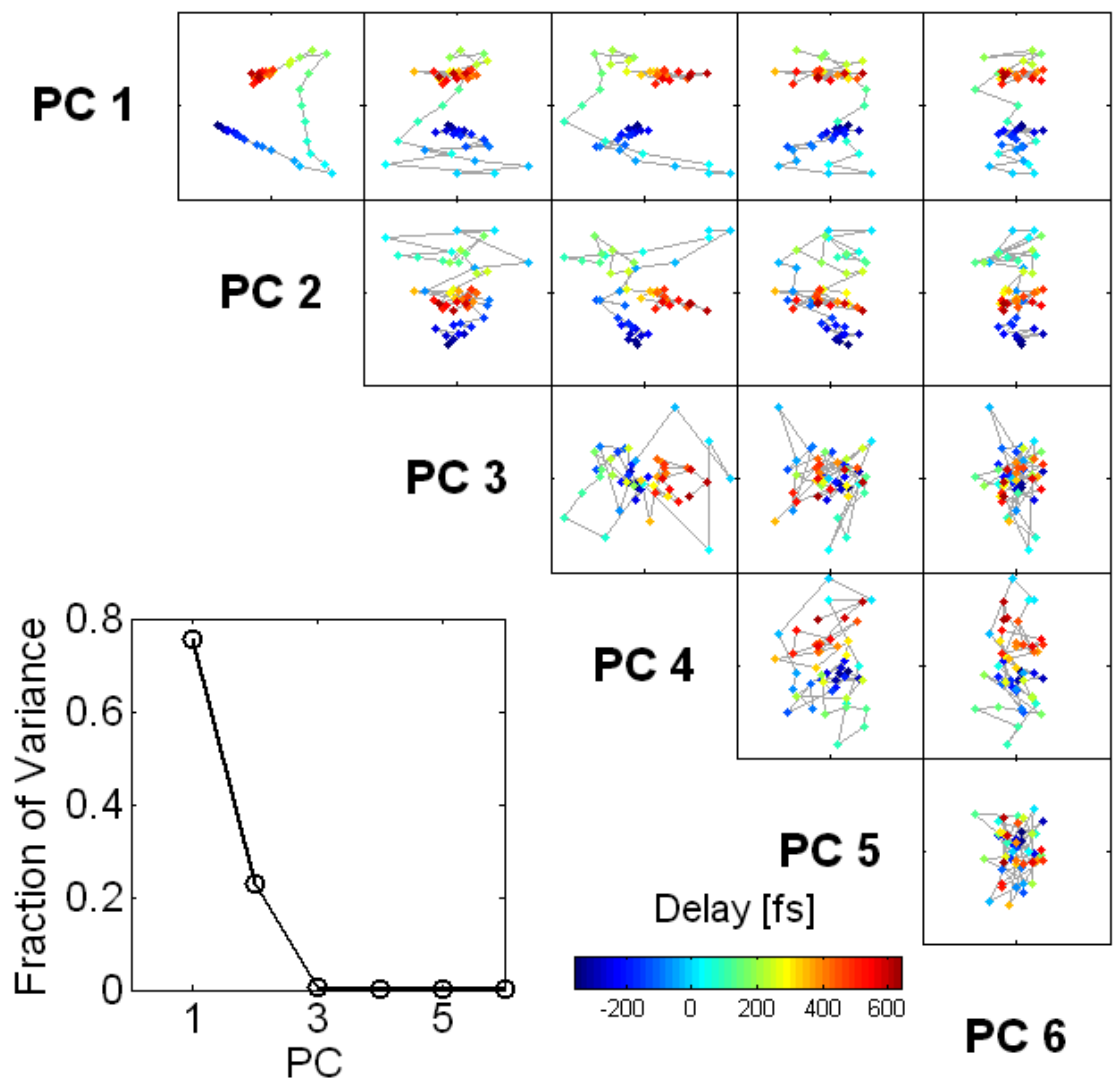

Figure 5: The TOF spectra projected onto different pairs of PCs (directions of greatest variance). Each point corresponds to a spectrum at a single delay. Lines connect spectra at adjacent delays, which are separated by 20 fs. Different pair plots are not shown at the same scale. (Inset) The fraction of variance accounted for by first six PCs. The fraction of variance is proportional to the corresponding eigenvalue of the covariance matrix.

Figure 5 shows the experimental mass spectra projected onto different pairs of principal components ordered according to decreasing variance. When projected onto the PC1-PC2 plane, the TOF spectra trace out a characteristic "cat face" (enlarged in figure 6) as a function of pump-probe delay. Multiple pump-probe experiments taken weeks apart and under slightly different experimental conditions all exhibit this characteristic cat shape and fall into the span of the same three endmembers[36]. The positions of the "ears" allow immediate identification of two characteristic times: the peak in the parent ion (left ear) and the H$^+$ signal maximum (right ear). The two maxima are separated by 180 fs, comparable to the findings of previous investigations [2, 7]. At long delay times, the product signal is constant.

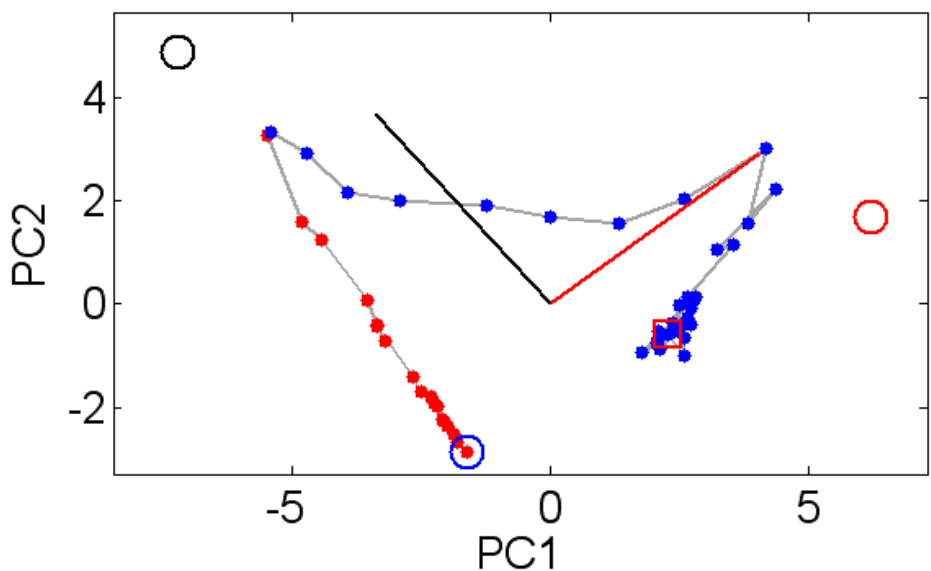

Figure 6: Experimental spectra and endmember spectra in the PC1-PC2 plane. The CHD (blue circle), CHD$^+$ (black circle), and HT (red circle) endmembers and the assumed 50:50 CHD:HT spectrum (red square) are also shown. The directions corresponding to an increase in H$^+$ (red line) and mass 80 (black line) are also shown.

When the spectra are viewed as vectors in a vector space, the trajectory is closely related to the underlying physical processes. For example, the straight edges of the data projection in PC space are significant features characteristic of conversion between two species when the total number of molecules is held constant. In our experiment, the number of molecules in the interaction region and the laser pulse energies are both approximately constant. If the cross-section for fragmenting a given endmember is constant as well (i.e., the probability of fragmenting a given species is constant), then the spectra resulting from the mixing of n endmembers fall into an n-1 dimensional space. For example, the spectra resulting from mixing two endmembers fall on a line connecting the two endmembers and for three endmembers the mixed spectra lie on the plane defined by the three endmembers. Thus straight lines are characteristic of the conversion of exactly one species into another and planarity is characteristic of conversions among 3 species.

**Endmember Spectrum Estimation**

The unmixing analysis can be taken one step further to estimate the pure spectra (position the endmembers in the plane of the first two PCs). First, a constant background due to the fragments created by the UV pulse alone is subtracted from the TOF spectra. The choice of this background affects the endmember spectra, but not the locations where they pierce the PC plane. In order to estimate the CHD endmember, we assume that only CHD is present at the earliest delay (when the probe comes ~400 fs before the pump) and we assign the earliest mass spectrum (see figure 4) to pure CHD.

In order to estimate the HT endmember spectrum, we assume that among molecules that absorbed one UV photon, the ratio of CHD:HT is 50:50 at the longest probe delays (~0.5 ps). This is the branching ratio estimated by theory [19]. (A CHD:HT branching ratio of 60:40 as measured in liquid phase [23] would place the HT endmember 25% farther away from the CHD endmember.) We also assume that the total number of molecules is independent of delay and that the fragmentation cross-section for the CHD endmember is the same at the earliest and longest delay. We then use the 50:50 CHD:HT spectrum to place the HT endmember as shown near the right ear; this endmember corresponds to having all excited molecules eventually isomerizing to HT. (Rather than relying on the spectrum at greatest delay to be 50:50 CHD:HT, we could have used purely geometric arguments to estimate the HT endmember. We could easily draw a small triangle enclosing the data and assign the vertex closest to the "right ear" to the HT endmember. The CHD:HT branching ratio at long delays would be roughly 50:50. However, this geometric approach fails if the data does not contain any adequately pure HT spectra, that is, spectra containing a high enough proportion of HT to permit reasonable estimation of the enclosing simplex.)

Following our conjecture that straight lines in the PC1-PC2 plot correspond to conversion between two endmembers, we posit that the $CHD^+$ endmember should roughly lie on the line connecting the earliest spectrum (CHD endmember) with the cross-correlation peak (left cat ear). We additionally require all time-of-flight mass peaks in the $CHD^+$ endmember to be positive. A further constraint is that all of the data should lie within the triangle defined by the three endmembers. These considerations result in the placement of the $CHD^+$ endmember spectrum as shown in figure 6.

In Fig. 6 we have also plotted the projections of the vectors corresponding to the $H^+$ direction (red line) and the parent ion direction (black line) onto the PC1-PC2 plane. This allows one to easily follow the behavior of different peaks during the delay scan. The parent ion count reaches a maximum near $T_0$ (near $CHD^+$) and the $H^+$ count reaches a maximum when the trajectory is closest to the HT endmember spectrum. The spectra at longer time delays contain more $H^+$ than the early time delays, reflecting a plateau in the count of hydrogen ions after the maximum. The $H^+$ and parent ion counts are shown in figure 10 for comparison.

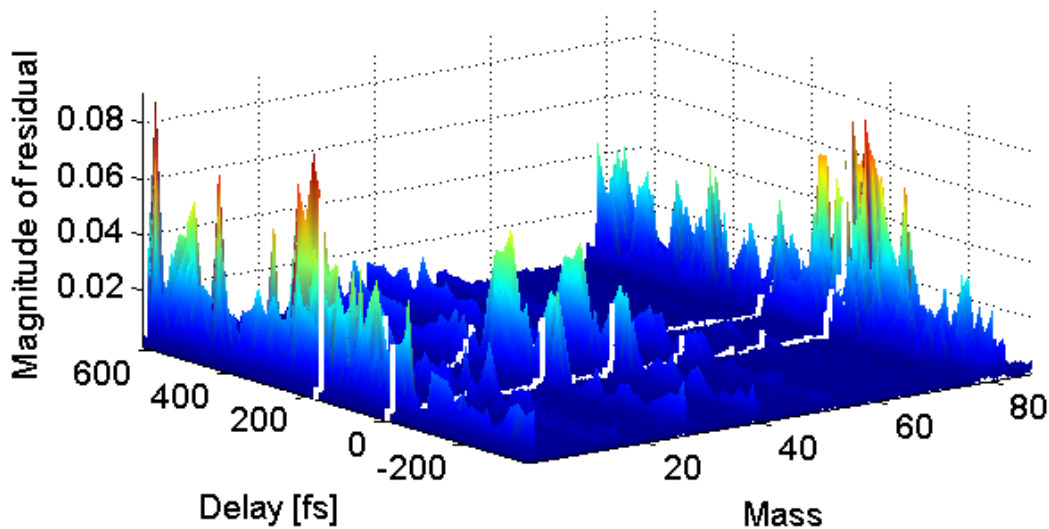

Figure 7: Magnitudes of residuals left after fitting the experimental spectra to three endmembers. The vertical scale is the same as in figure 4. The delays corresponding to the maximum yields of masses 1 and 80 are highlighted.

The residuals from fitting the experimental spectra to the simplex are shown in figure 7. Any choice of endmember spectra that spans the same space will fit the data equally well, and PCA gives a bound on the goodness-of-fit achievable. It is not possible to reduce the residuals by simply choosing the endmembers differently.

The endmember spectra are shown in figure 8. These spectra depend on the choice of constant background. The HT endmember spectrum is the only one with a significant contribution from $H^+$, and other light fragments including both $C^+$ and $C^{++}$ are also more common in the HT spectrum. The CHD endmember spectrum contains relatively few light fragments (methyl, mass 15, is a notable exception). The $CHD^+$ endmember spectrum is primarily composed of $CHD^+$ (mass 80) ions, as might be expected. We note, however, that even when the mass 1 and mass 80 ions are removed from the data set, the spectra still fall into a plane and trace out a triangular "cat face". In other words, it is possible to separate spectra from these three products solely on the basis of the intermediate mass fragments, even if the mass 1 and mass 80 components are neglected.

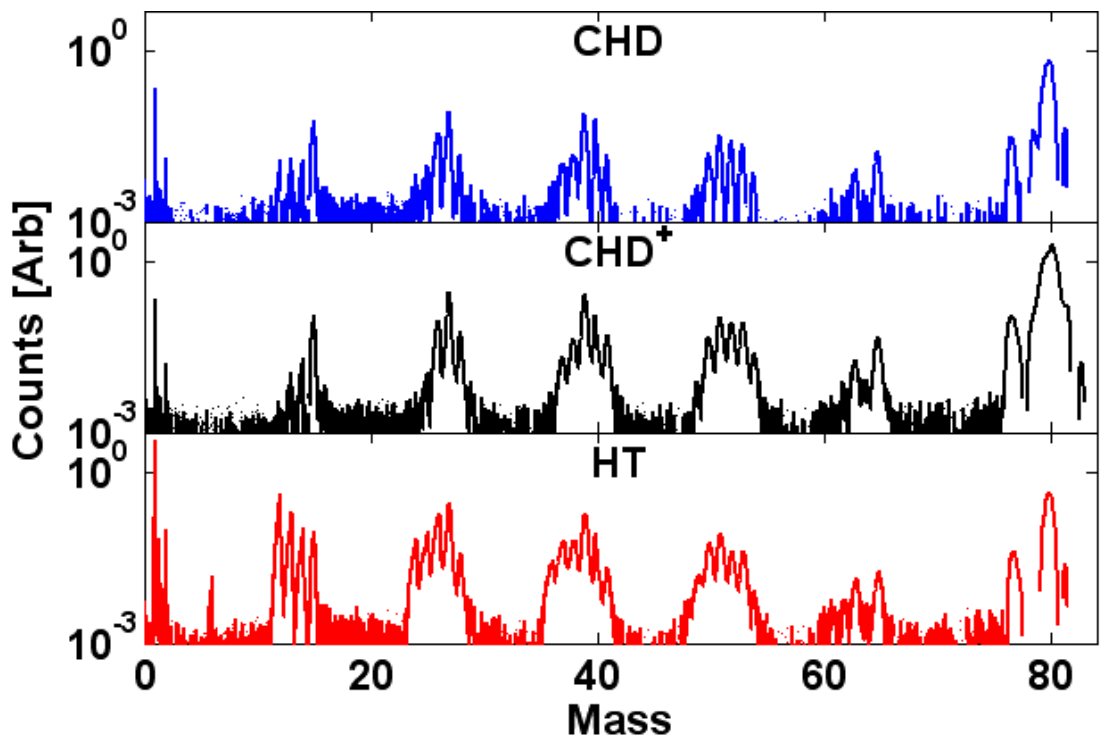

Figure 8: Endmember spectra determined by using the UV background as the origin plotted on a logarithmic scale.

The maximum in the mass 80 signal and the corresponding maximum in the $CHD^+$ endmember spectrum contribution at zero time delay are likely due to the temporal overlap of the pump and probe beams at this point. This is consistent with a multiphoton ionization pathway for CHD involving one UV photon together with multiple IR photons. In like manner, the maximum in $H^+$ near a delay of 200 fs suggests a peak in HT production, but it could be due instead to a transient increase in the photofragmentation cross section for HT as suggested by Fuss and Kotur [2, 7]. They point out that a transient resonance is suggested by the lateness of the $H^+$ maximum (200 fs) and the opportunity for resonances as the wavepacket accelerates downhill. The present analysis allows us to explore whether this apparent transient resonance is a new fragmentation channel, accompanied by its own distinct fragmentation pattern, or whether it is a transient increase in the contribution from the same HT endmember spectrum already identified.

At its maximum, the $H^+$ signal is approximately 50% higher than at long delays. If a distinct transient process is responsible for the increase, the experimental spectra near the HT maximum (right cat ear) should leave the simplex, i.e. the space enclosed by the endmembers. Figure 7 shows no transient increase in the residuals outside of the cat plane, which indicates that the data do not leave the simplex in the region of the resonance.

A transient enhancement of the same fragmentation route for HT will leave the cat ear in the original simplex, but generally bent out of the cat plane to indicate an enhanced HT spectrum. This is consistent with the data. The large constant background and noise in our measurement make it difficult to determine the precise geometry of any out-of-plane excursion (Fig. 7 and Fig. 9). Figure 9 shows the deviation from the plane in all dimensions. The deviation from planarity, taking into account all dimensions, is small compared to variation within the plane.

We identify $H^+$ as a signal of HT production. In particular, we find that the maximum in $H^+$ corresponds to a maximum in the entire HT endmember signal. One explanation for the constancy of the fragmentation patterns is that the resonance is enhancing the transfer of population to a state that is a precursor to fragmentation.

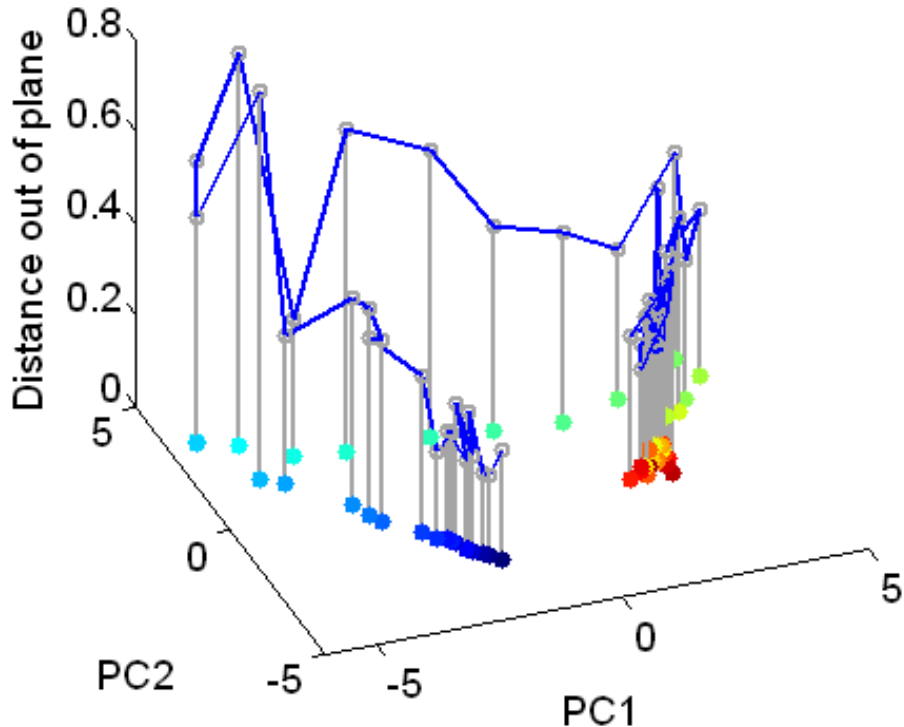

Figure 9: Residuals from fitting experimental spectra to the 2D affine space formed from the mean vector and the first two principal components.

**Proportion Determination**

At this point we have created a useful model but it suffers from proportion indeterminacy. In other words, we need to estimate the proportionality between endmember spectrum size (1-norm) and the number of endmember molecules in the interaction region (i.e., an estimate of the fragmentation cross-section for each endmember species). If we assume that the total number of molecules is constant and the fragmentation cross-section for each endmember is independent of delay, then for each mixed spectrum the endmember spectrum coefficients are equal to the proportions. These are the simplest assumptions consistent with the experiment.

The proportions are shown as a function of pump-probe delay in figure 10 (left). For comparison, the $H^+$ and parent ion signal are shown in figure 10 (right). The proportions given are for excited molecules that interact with the probe; the majority of the molecules are not excited and contribute a nearly constant pure CHD background. In other words, if we assume that 5% of the molecules are excited by the UV pulse, then the HT endmember location actually corresponds to spectra resulting from 5% HT and 95% CHD background (and likewise for the $CHD^+$ endmember). In the PC plane, the spectra corresponding to 100% HT would then lie 20 times farther away than the HT endmember shown. The CHD endmember does not suffer from this problem because the unknown background is also all CHD (and the constant background due the UV pump is subtracted).

The noise on the $H^+$ lineout is clearly reflected in the HT endmember proportion. This noise is largely due to fluctuations in the number of HT molecules being detected. We believe that the fluctuations are mostly due to UV pulse energy fluctuations.

The experimental mass spectra are well-modeled by three fragmentation patterns. The HT proportion closely follows the $H^+$ signal. In particular, the maximum in the $H^+$ peak is part of an overall increase in the HT fragmentation pattern. The transient increase in the HT proportion around ~200 fs delay is indicative of an increase in the number of HT molecules being detected (perhaps due to a resonance, as discussed above). Interestingly, the characteristic HT fragmentation pattern is unchanged during the transient increase. The $CHD^+$ proportion follows reaches a maximum near zero delay.

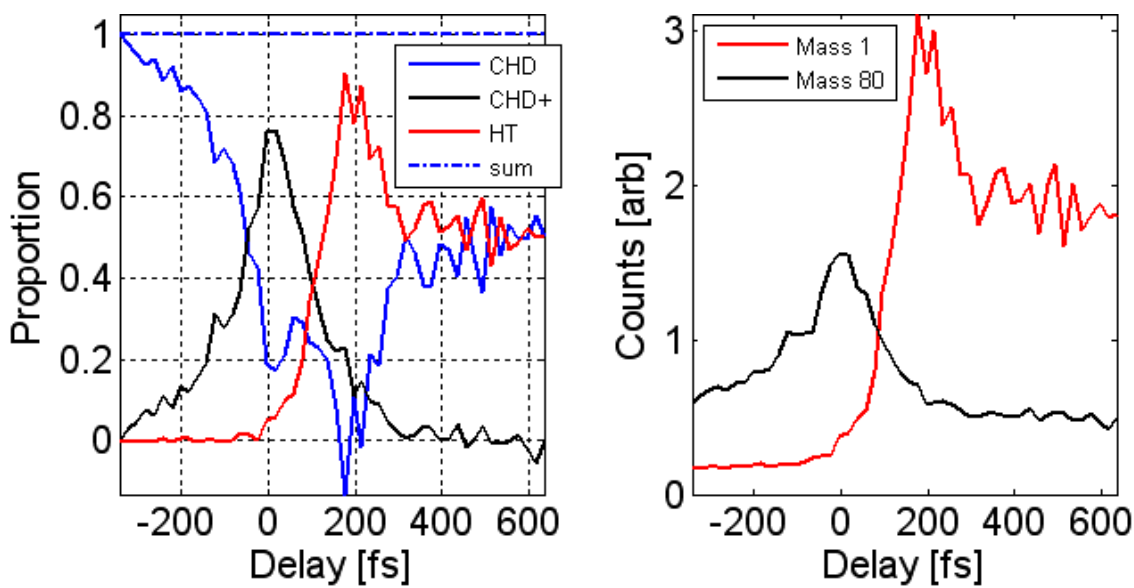

Figure 10: (left) Endmember proportions vs. delay. The spectra were projected into the PC plane before calculating the proportions, which forces the proportions to sum to one. The CHD:HT ratio at long times is fixed at 50:50. (right) The parent peak (black) and $H^+$ peak (red) vs. delay. A constant background due to the UV pulse has been subtracted.

**V. Conclusion**

We have used a simplex-based spectral unmixing technique to analyze TOF spectra collected during a pump-probe experiment that monitored the photoinitiated isomerization of CHD into HT. The technique allows the determination of the number of endmember spectra (pure spectra) necessary to model the data, estimation of endmember spectra, and estimation of their proportions in the experimental spectra (unmixing). We determined that transient TOF spectra from the CHD-HT ring-opening reaction could be fit with three endmembers. Using a combination of physical knowledge and the simplex-based spectral unmixing technique, we estimated the endmember spectra and assigned them to CHD, $CHD^+$, and HT. We find that $H^+$ is due almost entirely to HT; in particular the increase in $H^+$ around 180 fs delay is part of an overall increase in the HT fragmentation pattern. We determined that the three fragmentation patterns represented by the three endmembers were sufficient to interpret the transient increase in the $H^+$ yield at this delay.

The reduction of the entire data set to mixtures of only three parent spectra is noteworthy, since the dynamics of the CHD-HT ring opening reaction is believed to involve nonradiative transitions between two excited states and a conical intersection to the ground state of CHD or HT. In particular, we find large spectral differences between CHD and HT but no discernible differences between the spectra of ground state CHD and vibrationally excited CHD. It is possible that the fragmentation pattern is largely determined by more general geometrical features, such as whether the ring is open or closed.

We have presented a simple interpretation of the data based on a consideration of the first two principal components. Given sufficient signal to background ratios, smaller principal components may be employed to identify further, distinct endmember spectra associated with either specific regions on the potential energy surface (PES) or distinct fragmentation channels. Examination of PC3 and PC4 displayed in figure 5 suggest that there might be additional channels.

The unmixing technique presented is widely applicable. Determination of the number of fragmentation patterns involved is vital to avoid underfitting the data and is easily accomplished using a dimension-reduction technique such as principal component analysis. The ability to estimate endmember spectra is especially useful when pure spectra are not available (for example, unstable species or spectra corresponding to certain regions on a PES). A particular strength of the method is that, when viewed as points in a vector space, the geometry of the spectra is closely related to underlying physical processes, such as conversion between species or increases in fragmentation cross sections. Low-dimensional models of experimental spectra are easily visualized and important features, such as the mixing of two endmembers, are readily identified. Finally, it is easy to create maps revealing how the spectra change as a function of experimental parameters, such as pump-probe delay.

**Acknowledgements:** This work was supported by the National Science Foundation.